\documentclass[journal]{IEEEtran}

\ifCLASSINFOpdf
\else
   \usepackage[dvips]{graphicx}
\fi
\usepackage{url}
\usepackage[cmex10]{amsmath}
\usepackage{amssymb}
\usepackage{mathrsfs}
\usepackage{verbatim}
\usepackage{graphicx}
\usepackage{epstopdf}
\usepackage{cite}
\usepackage[justification=justified]{caption}
\usepackage{ifthen}

\newboolean{singleColumn}
\setboolean{singleColumn}{false}

\begin{document}

\title{Equiripple MIMO Beampattern Synthesis using Chebyshev Approximation}

\author{David A. Hague, \IEEEmembership{Member, IEEE} and David G. Felton, \IEEEmembership{Student Member, IEEE}
\thanks{Submitted March 18, 2024.  This research was funded by the Naval Undersea Warfare Center's In-House Laboratory Independent Research (ILIR) program from the Office of Naval Research (ONR) under N0001424WX00177.}
\thanks{D. A. Hague is with the Naval Undersea Warfare Center, Newport, RI 02841 USA (e-mail: david.a.hague.civ@us.navy.mil).}
\thanks{D. G. Felton is with the University of Kansas, Lawrence, KS USA.  (e-mail: davidfelton42@ku.edu).}}

\markboth{IEEE Signal Processing Letters,~Vol.~xx, No.~x, January~2025}%
{Hague \MakeLowercase{\textit{et al.}}: MIMO Beampatterns using Chebyshev Approximation}

\maketitle

\begin{abstract}
This letter presents a method for synthesizing equiripple MIMO transmit beampatterns using Chebyshev approximation.  The MIMO beampattern is represented as a non-negative real-valued trigonometric polynomial where the $\ell^{\text{th}}$ order polynomial coefficient is the sum of the $\ell^{\text{th}}$ order diagonal of the waveform correlation matrix.  The optimal coefficients for a given equiripple beampattern design is then posed as a Chebyshev approximation problem which is efficiently solved using the Parks-McClellan algorithm from optimal Finite Impulse Response (FIR) filter design theory.  The unique advantage of this synthesis method is that it provides a closed form method to generating MIMO correlation matrices that realize the desired equiripple beampattern.  This correspondingly facilitates the design of waveform sets that closely approximate those correlation matrices.  This method is demonstrated via two illustrative design examples; the first using traditional partial signal correlation methods and the second using transmit beamspace processing.  Both examples realize equiripple beampatterns using constant envelope and spectrally compact waveform sets.  
\end{abstract}


%
\IEEEpeerreviewmaketitle

\section{Introduction}
\label{sec:Intro}
%
%
%
%
\IEEEPARstart{M}{ultiple}-Input Multiple-Output (MIMO) arrays have seen extensive use in communications and radar applications \cite{JianLiBookI}.  A colocated MIMO array is a generalization of the standard phased-array that transmits a unique waveform on each element.  This increases the number of degrees of freedom in the transmit array design which facilitates synthesizing novel transmit beampattern shapes.  The structure of the MIMO correlation matrix, whose entries are the inner products between each of the waveforms in the waveform set, determines the shape of the resulting transmit beampattern of the MIMO array.  The MIMO beampattern synthesis problem is a two step process involving (1) finding a correlation matrix that realizes a close fit to a desired beampattern shape \cite{MIMO_SanAntonio, MIMO_Li_I, Sajid_MIMO_Matrix_1, Visa_MIMO_Matrix} and (2) designing waveform sets that at a minimum approximate that correlation matrix \cite{MIMO_SanAntonio, MIMO_Waveform_Sajid, jianLiII, MIMO_Li_II, MIMO_Li_III}.  There exists a rich literature for both problems \cite{RangaswamyI, PalomarI, Prabhu_Babu_I, Tang_Spectra, Blum_MIMO_Waveform_1, Monga_MIMO, Cui_MIMO, Himed_MIMO_Waveform} that continues to grow \cite{SPL_1, SPL_2}.

The majority of the efforts in the literature focus on designing beampatterns with a nominally flat passband region, a narrow transition band region, and a stopband region where almost no power is transmitted.  As a consequence of the Gibbs phenonmenon, the beampattern's passband \& stopband regions possess ripples.  The magnitude of these ripples appears to be controlled by the width of the transition band region in a manner that is strongly similar to that of Finite Impulse Response (FIR) filter design \cite{MIMO_IS_FIR}.  In fact, the development of MIMO beampattern synthesis methods has largely followed a parrallel path to that of optimal FIR filter design.  Work by \cite{MIMO_SanAntonio, MIMO_Li_II} developed a series of algorithms that produced minimum mean-square error (MMSE) and minimax fits to desired beampatterns which are analogous to the efforts of \cite{Rabinar_FIR_Opt, Kaiser_Filter} that developed numerical optimization techniques for FIR filters.  Work by \cite{Fourier_MIMO} synthesized MIMO beampatterns using a Fourier sampling technique that is largely reminiscent of the work by Rabiner \emph{et. al.} \cite{Rabiner_Fourier_Sampling} for digital filter design.  There are also several efforts that focus on beampattern synthesis with passband and stopband ripple control \cite{MIMO_TBP_Hua, Augusto_PSL, MIMO_Li_Ripple, MIMO_Lp_Trans} which mirrors some of the early attempts at equiripple FIR filter design \cite{FIR_Equiripple}.  Interestingly, \cite{MIMO_TBP_Hua} noted a relationship between the ripple levels, the length of the array, and the transition bandwidth of the beampattern that closely resembles the equiripple approximations developed by James Kaiser \cite{parksMcClellan_Main}. 

What appears to be missing from the MIMO literature is a method that directly synthesizes equiripple MIMO beampatterns in a manner similar to the optimal FIR filter design techniques developed by Parks and McClellan \cite{parksMcClellan_Main, parksMcClellan_Orig}.  This letter takes a first step in this direction by introducing a method of equiripple MIMO beampattern synthesis using Chebyshev approximation.  We show that an even-symmetric MIMO beampattern is expressed as a nonnegative trigonometric polynomial represented as a finite Fourier cosine series \cite{Dumitrescu}.  This representation is identical to the frequency response of Type-I FIR filters \cite{parksMcClellan_Main}.  The $\ell^{\text{th}}$ order Fourier coefficient is equal to the sum of the $\ell^{\text{th}}$ order diagonal of the MIMO waveform correlation matrix and are readily found for any equiripple MIMO beampattern via the Parks-McClellan algorithm \cite{parksMcClellan_Main, parksMcClellan_Orig}.  This method also provides insight into the structure of the MIMO waveform correlation matrices that realize equiripple beampatterns.  This in turn facilitates developing simple algorithms to synthesize constant-envelope/spectrally-compact waveform sets that realize these matrices.  

\section{MIMO Signal Model and Chebyshev Approximation}
\label{sec:MIMO}
This section describes the MIMO waveform signal model and introduces the Multi-Tone Sinusoidal Frequency Modulated (MTSFM) waveform model used to realize MIMO beampatterns in this letter.  The model assumes a Uniform Linear Array (ULA) with $M$ elements and half-wavelength inter-element spacing $d=\lambda/2$.

\subsection{The MIMO Waveform Set}
Each element of a colocated MIMO ULA transmits a unique basebanded waveform expressed in discrete time as $x_m\left[n\right]$ where $m=1,\dots,M$ and $n=0,\dots,N-1$.  Each of the $M$ waveforms in the waveform set possess $N$ discrete time samples and is expressed as
\begin{equation}
x_m\left[n\right] = \sqrt{\dfrac{E}{MN}}e^{j\varphi_m\left[n\right]}
\label{eq:waveformSet}
\end{equation}
where $\varphi_m\left[n\right]$ is the $m^{\text{th}}$ waveform's instantaneous phase and the $\sqrt{E/MN}$ term normalizes each waveform to have a total energy of $E/M$.  We express \eqref{eq:waveformSet} in vector form as $\mathbf{x}_m \in \mathcal{C}^{1\times N}$.  This paper utilizes the MTSFM waveform model in \cite{Hague_AES} whose phase modulation function is a finite Fourier sine series
\begin{equation}
\varphi_m\left[n\right] = \sum_{p=1}^P \alpha_{m,p} \sin \left(\dfrac{2\pi p n}{N}\right).
\label{eq:MTSFM}
\end{equation}
where $\alpha_{m,p}$ are the $P$ Fourier coefficients for each of the $M$ waveforms in the set representing a discrete set of $MP$ coefficients.  Note that \eqref{eq:MTSFM} can also include cosine harmonics in the phase modulation function.  However, this letter focuses solely on MTSFM waveforms with odd-symmetric phase modulation functions for mathematical simplicity as in \cite{Hague_Asilomar_2024}.

\subsection{The MIMO Beampattern and Correlation Matrix}
\label{subsec:ulsModel}
The waveform set is represented in matrix form as 
\begin{equation}
\mathbf{X}= \left[\mathbf{x}_1^T, \mathbf{x}_2^T, \dots, \mathbf{x}_M^T \right]^T \in \mathcal{C}^{M\times N}.
\end{equation}
The narrowband MIMO transmit beampattern is a normalized power density across the spatial angle $u=\pi\sin\theta$
\begin{equation}
P\left(u\right) = \mathbf{a}^H\left(u\right) \mathbf{R} \mathbf{a}\left(u\right) \geq 0,~-\pi\leq u \leq \pi
\label{eq:mimoBeamPattern}
\end{equation}
where $\mathbf{a}\left(u\right)$ is the $M\times 1$ transmit array steering vector 
\begin{equation}
\mathbf{a}\left(u\right) = \left[1,~e^{j u},~\dots,~e^{j\left(M-1\right)u} \right]^{\text{T}}.
\label{eq:mimoSteerVec}
\end{equation} 
The matrix $\mathbf{R} \in \mathcal{C}^{M\times M}$ is the MIMO waveform correlation matrix of the waveform set whose entries represent the inner products between all waveforms in the set \cite{MIMO_SanAntonio}
\begin{equation}
\mathbf{R} = \mathbf{X}\mathbf{X}^H.
\label{eq:mimoCorrMat}
\end{equation}
There are several conditions governing the MIMO correlation matrix $\mathbf{R}$.  The primary condition on \eqref{eq:mimoCorrMat}, which directly follows from \eqref{eq:mimoBeamPattern}, is that it must be positive semi-definite, denoted as $\mathbf{R} \succeq 0$.  There is often an additional constraint that each element transmits equal power and thus the diagonals of $\mathbf{R}$ are all $E/M$.  However, this constraint can be relaxed to nonuniform power across elements as long as $\text{tr}\{\mathbf{X}\mathbf{X}^H \} = E$.

As has been noted in many prior efforts \cite{MIMO_SanAntonio, MIMO_Li_II, MIMO_Waveform_Sajid}, the design of waveform sets that realize a desired correlation matrix is a challenging problem.  Transmit Beamspace Processing (TBP) \cite{MIMO_TBP_Hassanien} was proposed as a more efficient means to address this waveform design problem.  The TBP technique introduces a weighting matrix to a set of orthogonal waveforms \cite{MIMO_TBP_Hua, MIMO_TBP_Hua_ICASSP}.  The weighted transmit waveform set is expressed as 
\begin{equation}
\tilde{\mathbf{X}} = \sum_{m=1}^M \mathbf{w}_m \mathbf{x}_m = \mathbf{W}\mathbf{X}
\label{eq:TBP}
\end{equation} 
where $\mathbf{W}$ is the $M\times M$ weights matrix whose $m^{\text{th}}$ column is $\mathbf{w}_m$.  The MIMO beampattern using \eqref{eq:TBP} is now
\begin{equation}
P\left(u\right) = \mathbf{a}^H\left(u\right) \mathbf{R_W} \mathbf{a}\left(u\right) = \mathbf{a}^H\left(u\right) \mathbf{W} \mathbf{W}^H \mathbf{a}\left(u\right)
\end{equation}
where $\mathbf{R_W}=\mathbf{W} \mathbf{X} \mathbf{X}^H \mathbf{W}^H=\mathbf{W}\mathbf{W}^H$ since the waveform set $\mathbf{X}$ is assumed to be orthogonal.  The most direct way to generate $\mathbf{W}$ for a given $\mathbf{R_W}$ is based on the eigen-decomposition of $\mathbf{R_W}$ \cite{MIMO_TBP_Hua_ICASSP}
\begin{equation}
\mathbf{W} = \mathbf{U}\sqrt{\mathbf{\Lambda}}
\end{equation}
where $\mathbf{U}$ and $\mathbf{\Lambda}$ are the matrix of eigenvectors and diagonal matrix of eigenvalues respectively.

\section{Equiripple MIMO Beampattern Design}
\label{sec:mimoBeam}
This section poses the equiripple MIMO beampattern design problem as a non-negative trigonometric polynomial whose coefficients can be found using Chebyshev approximation via the Parks-McClellan algorithm.  This design method is then demonstrated using two design examples. 

\subsection{Problem Definition}
\label{subsec:problemDef}
\begin{figure}[ht]
\centering
\ifthenelse {\boolean{singleColumn}}
 {\includegraphics[width=1.0\textwidth]{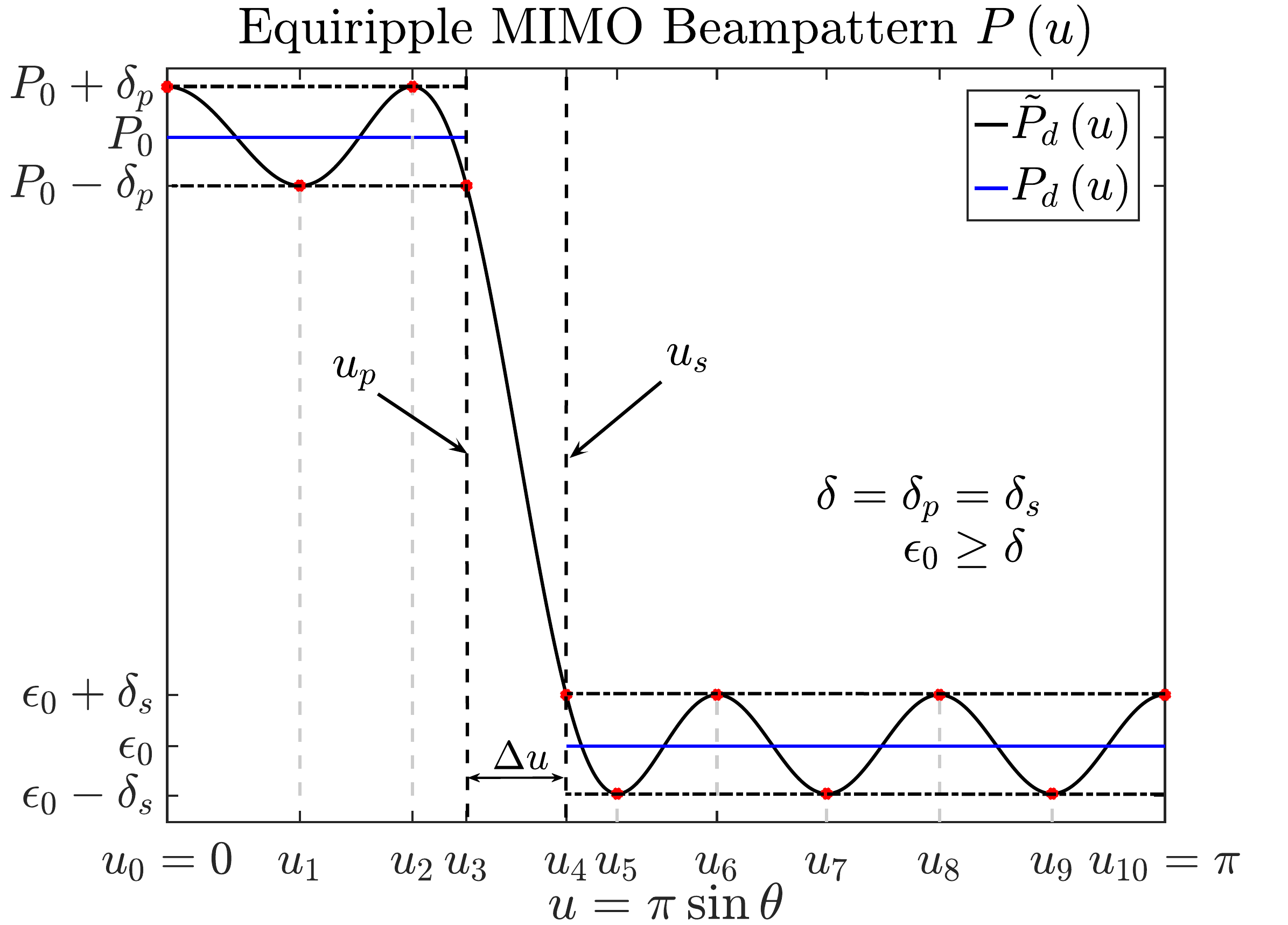}}
 {\includegraphics[width=0.5\textwidth]{Figure1.pdf}}
\caption{Specification of the equiripple MIMO beampattern synthesis problem with desired beampattern $P_d\left(u\right)$ and its equiripple approximation $\tilde{P}_d\left(u\right)$.  The stopband region height is set to $\epsilon_0 \geq \delta$ to ensure $\tilde{P}_d\left(u\right) \geq 0$ for all $u$.}
\label{fig:SPS_Figure_1}
\end{figure}

This letter specifically focuses on MIMO beampatterns that are even-symmetric in angle for mathematical simplicity.  However, we note that this method can be readily extended to arbitrary symmetry.  As shown in Figure \ref{fig:SPS_Figure_1}, the equiripple MIMO beampattern synthesis problem is essentially the same as the equiripple FIR filter design problem.  There is a  passband region $|u| \leq u_p$ across which the beampattern is flat at some power level $P_0$ with ripple $\delta_p=\delta$.  There is also a well defined transition band $\Delta u = u_s-u_p$.  What is different is that the stopband region $|u| \geq u_s$ has some non-zero level $\epsilon_0$ that must be greater than the stopband ripple $\delta_s=\delta_p=\delta$.  This ensures that $P\left(u\right)$ is nonnegative for all $u$.  Correspondingly, setting $\epsilon_0 = \delta$ establishes the minimum achievable peak sidelobe height of the beampattern to be $2\delta$.  

In order to utilize the Parks-McClellan algorithm for equiripple MIMO beampattern synthesis, the MIMO beampattern representation in \eqref{eq:mimoBeamPattern} must be expressed in the appropriate form that can leverage Chebyshev approximation.  Inserting \eqref{eq:mimoSteerVec} into \eqref{eq:mimoBeamPattern} and expanding terms results in the expression \cite{Dumitrescu}
\begin{align}
P\left(u\right) = \sum_{m=1}^M \sum_{m'=1}^M \mathbf{R}_{m,m'} e^{-j\left(m-m'\right)u}
\label{eq:mimoBeamPattern1}
\end{align}
where $\mathbf{R}_{m,m'}$ are the entries of the waveform set's correlation matrix as defined in \eqref{eq:mimoCorrMat}. Letting $\ell = m-m'$ and rearranging the order of the summation simplifies \eqref{eq:mimoBeamPattern1} to
\begin{equation}
P\left(u\right) = \sum_{\ell=0}^{M-1}\sum_{m=1}^{M-\ell}\mathbf{R}_{m, m+\ell}e^{j\ell u}.
\label{eq:mimoBeamPattern2}
\end{equation}
Lastly, an even-symmetric beampattern results in a real-symmetric correlation matrix (i.e., $\mathbf{R}_{m,m'} = \mathbf{R}_{m',m}$).  Using Euler's identity, \eqref{eq:mimoBeamPattern2} then simplifies to
\begin{equation}
P\left(u\right) = r_0 + 2\sum_{\ell=1}^{M-1} r_{\ell}\cos\left(\ell u\right),~r_{\ell} = \sum_{m=0}^{M-\ell}\mathbf{R}_{m, m+\ell}.
\label{eq:mimoBeamPattern3}
\end{equation}
The expression for the MIMO beampattern in \eqref{eq:mimoBeamPattern3} essentially possesses the form of a Type-I FIR filter frequency response \cite{parksMcClellan_Main}.  Finding the coefficients $r_{\ell}$ to synthesize an equiripple beampattern can then be achieved using the Parks-McClellan algorithm \cite{parksMcClellan_Orig}.  Choosing the necessary number of sensors $M$ to realize an equiripple beampattern shape with a desired $\Delta u$ and $\delta$ directly follows from a modified version of Kaiser's formula for FIR filters \cite{parksMcClellan_Main}
\begin{equation}
M \approx \dfrac{-20\log_{10}\left(\delta\right) -13}{14.6 \Delta u}+1.
\label{eq:Kaiser}
\end{equation}
Nonnegative trigonometric polynomials have been used in the design of FIR filters \cite{Boyd_FIR_I, Dumitrescu_FIR} and appears to have only been briefly explored for MIMO beampattern synthesis \cite{IEEE_Trans_Poly, Visa_MIMO_Poly}.  To the best of the authors' knowledge, applying Chebyshev approximation to \eqref{eq:mimoBeamPattern3} to synthesize equiripple MIMO beampatterns appears to be novel. 

The Chebyshev coefficients in \eqref{eq:mimoBeamPattern3} not only produce the desired equiripple MIMO beampattern, but it also provides insight into the structure of the waveform correlation matrix that realizes it.  Perhaps the simplest matrix construction is a Toeplitz structured matrix
\begin{equation}
\mathbf{R} = \left[\begin{matrix}
			  \tilde{r}_0 		& \tilde{r}_1 		& \tilde{r}_2 		& \dots 		& \tilde{r}_{M-1} 	\\
			  \tilde{r}_1 		& \tilde{r}_0 		& \tilde{r}_1 		& \dots 		& \tilde{r}_{M-2} 	\\
			  \tilde{r}_2 		& \tilde{r}_1 		& \tilde{r}_0 		& \dots 		& \tilde{r}_{M-3} 	\\ 
			  \vdots       		& \vdots       		& \vdots       		& \ddots 	& \vdots            		\\
			  \tilde{r}_{M-1}  & \tilde{r}_{M-2} & \tilde{r}_{M-3} & \dots 		& \tilde{r}_0
			  \end{matrix}\right]
\label{eq:ToeplitzMat}
\end{equation}
where $\tilde{r}_{\ell}=r_{\ell}/\left(M-\ell\right)$.  Indeed, Toeplitz structured matrices have been proposed in other MIMO beampattern synthesis problems \cite{MIMO_SanAntonio, Fourier_MIMO}.  However, as noted in \cite{Dumitrescu}, not all Toeplitz matrices of this kind are guaranteed to be positive semi-definite which violates a foundational property of MIMO correlation matrices.  Fortunately, from \eqref{eq:mimoBeamPattern3}, there's essentially an unlimited number of correlation matrices that realize any MIMO beampattern in general.  The only requirement for them is that the sum of their $\ell^{\text{th}}$ order diagonals equal the Fourier coefficients.  Additionally, for TBP, there exists at most $2^{M-1}-1$ beamforming vectors with the same beampattern \cite{Hassanien_2015}.  This flexibility in the structure of the correlation matrices and beamforming vectors greatly simplifies the synthesis of these matrices and the waveform sets that realize them.  
  
\subsection{Some Illustrative Design Examples}
\label{subsec:results}
This section provides two illustrative design examples to demonstrate the equiripple MIMO beampattern synthesis method using Chebyshev approximation.  For both design examples, MTSFM waveforms with a time-bandwidth product of $T\Delta f = 64$ are utilized as in \cite{Hague_Asilomar_2024} where $T$ is the waveform duration and $\Delta f$ is the waveform's swept bandwidth.  The number of harmonics $P$ in each MTSFM waveform's instantaneous phase \eqref{eq:MTSFM} were chosen to be $P=\left\lceil T \Delta f/2\right\rceil = 32$ to ensure a compact spectral shape \cite{Hague_AES, Hague_Asilomar_2024}.  The waveform time-series are sampled at a rate $f_s=5\Delta f$ and therefore possess $N=Tf_s + 1=5 T \Delta f + 1 = 321$ samples each.  

\subsubsection{ULA with $M=10$ Elements with Unit Elemental Power}
\label{subsubsec:exampleI}
The first design example utilizes a ULA with $M=10$ elements each transmitting equal power.  The desired beampattern has unity gain (i.e $P_0=1$) in the passband defined as $|u| \leq u_p =0.2\pi$.  The stopband is defined as $u_s =0.4\pi \leq |u| \leq \pi$ with a constant power level of $\epsilon_0 = 0.05$.  Equation \eqref{eq:Kaiser} dictates that the minimum ripple achievable is $\delta \approx 0.0112$.  The peak sidelobe level in the stopband is thus $\epsilon_0+\delta \approx 0.0612$ or $-12.1$ dB.  The Fourier cosine coefficients $r_{\ell}$ from \eqref{eq:mimoBeamPattern3} for this equiripple beampattern design problem are found using the Parks-McClellelan algorithm.  The corresponding correlation matrix resulting from \eqref{eq:mimoBeamPattern3} can be represented in the Toeplitz form according to \eqref{eq:ToeplitzMat} since it is positive semi-definite.

The Fourier cosine coefficients of the MTSFM waveform set that realizes this equiripple beampattern must match those found using the Parks-McClellan agorithm.  This is achieved by solving the constrained waveform optimization problem
\begin{multline}
\underset{\alpha_{m,p}}{\text{min}}\|r_{\ell}-\hat{r}_{\ell}\{\alpha_{m,p}\} \|_2^2 \\ \text{s.t.~} \beta_{rms}^2\left(\{\alpha_{m,p}^{\left(i\right)}\}\right) \in \left(1\pm\mu\right)\beta_{rms}^2\left(\{\alpha_{m,p}^{\left(0\right)}\}\right)
\label{eq:Problem1}
\end{multline}  
where $r_{\ell}$ are the Fourier cosine coefficients found using the Parks-McClellan algorithm, $\hat{r}_{\ell}\{\alpha_{m,p}\}$ are the coefficients resulting from the respective diagonals of the MTSFM waveform set's correlation matrix $\mathbf{R}\{\alpha_{m,p}\}$, $\beta_{rms}^2\left(\{\alpha_{m,p}^{\left(i\right)}\}\right)$ is the Root Mean Square (RMS) bandwidth \cite{Cohen} of each MTSFM waveform in the set at the $i^{\text{th}}$ iteration of the routine, and $0<\mu<1$ is a unitless constant.  The RMS bandwidth constraint, as discussed in \cite{Hague_AES}, ensures that each MTSFM waveform's spectral extent stays within some tolerance $\mu$ of the initial waveforms passed to \eqref{eq:Problem1} thus preserving its compact spectral shape \cite{Hague_Asilomar_2024}.  The optimization routine \eqref{eq:Problem1} is solved using MATLAB's optimization toolbox \cite{Matlab}.  

\begin{figure}[ht]
\centering
\ifthenelse {\boolean{singleColumn}}
 {\includegraphics[width=1.0\textwidth]{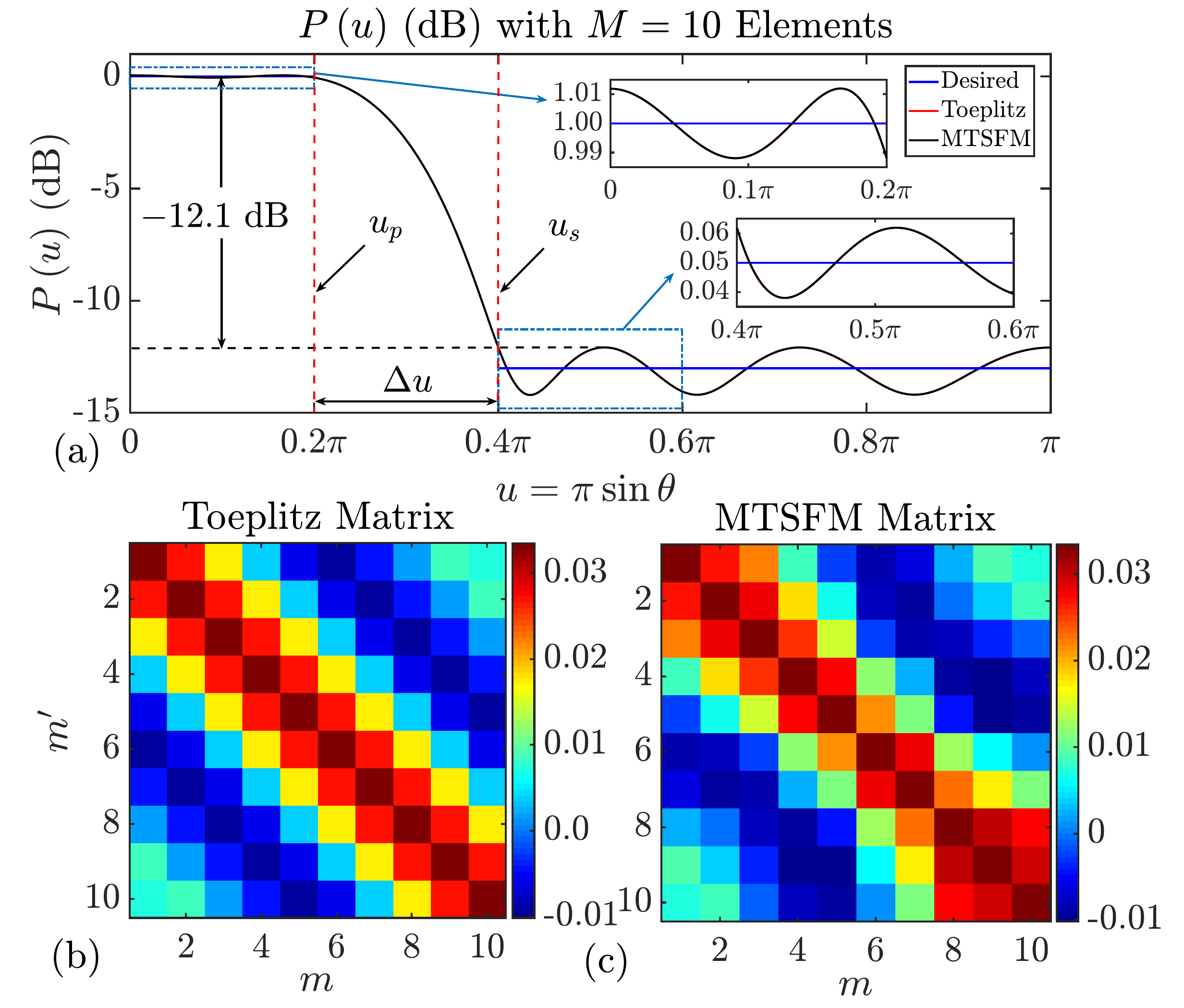}}
 {\includegraphics[width=0.5\textwidth]{Figure2.pdf}}
\caption{Equiripple MIMO beampatterns (a) synthesized from the Toeplitz structured correlation matrix (b) and the MTSFM waveform set correlation matrix (c).  Both matrices realize the desired equiripple beampattern design while their off-diagonal elements are noticeably different.}
\label{fig:SPS_Figure_2}
\end{figure}

The resulting equiripple beampatterns from the Toeplitz structured correlation matrix and the MTSFM waveform set are shown in panel (a) of Figure \ref{fig:SPS_Figure_2}.  Also shown in the figure are the Toeplitz and MTSFM correlation matrices that realize this equiripple beampattern.  As can be seen in the figure, both equiripple beampatterns are identical to machine precision with a passband/stopband ripple of $\delta\approx 0.0112$ resulting in a peak sidelobe of roughly $-12.1$ dB.  Note that while both beampatterns are identical, their respective correlation matrices possess different values for their off-diagonal entries.  However, their summations along the $\ell^{\text{th}}$ order diagonals are identical, which is a consequence of \eqref{eq:mimoBeamPattern3}.

\subsubsection{ULA with $M=20$ Elements using TBP}
\label{subsubsec:exampleII}
The second design example utilizes a ULA with $M=20$ elements.  Again, the desired beampattern has a unity gain $P_0=1$ in the passband defined as $|u| \leq u_p = 0.2\pi$.  The stopband is defined as $u_s = 0.4\pi \leq |u| \leq \pi$ with a constant power level of $\epsilon_0 = \delta = 0.000339$ derived via \eqref{eq:Kaiser}.  The resulting peak sidelobe level in the stopband is thus $\epsilon_0+\delta = 0.000678$ or $-31.7$ dB.  Once again, the coefficients $r_{\ell}$  from \eqref{eq:mimoBeamPattern3} are found using the Parks-McClellelan algorithm.  However, the resulting Toeplitz structured correlation matrix is not positive semi-definite.  Heuristically, the authors found that for beampattern designs with equal element power and sidelobe levels below $\approx -13.2$ dB, the Toeplitz structured matrix was no longer positive semi-definite. Therefore, this example will use TBP with nonuniformly weighted elements that transmit orthogonal MTSFM waveforms to realize the equiripple MIMO beampattern.  Finding a positive semi-definite correlation matrix that satisfies \eqref{eq:mimoBeamPattern3} is efficiently solved using the algorithm specified in (2.14) of \cite{Dumitrescu} which also provides an efficient implementation using the CVX software package \cite{cvx}.  As was noted in \cite{MIMO_TBP_Hua_ICASSP}, even small but nonzero correlation values between any two waveforms in the set can introduce perturbations in the TBP process that degrades the ideal MIMO beampattern shape.  To minimize this effect, the MTSFM waveform set is optimized to be as nearly orthogonal as is practical with finite time-bandwidth product waveforms.  This is achieved by solving the following constrained optimization problem
\begin{multline}
\underset{\alpha_{m,p}}{\text{min}}\left(\frac{E}{M}\right)\Bigl\|\mathbf{R}\left\{\alpha_{m,p}\right\}-\mathbf{I_{M\times M}} \Bigr\|_F^2 \\ \text{s.t.~} \beta_{rms}^2\left(\{\alpha_{m,p}^{\left(i\right)}\}\right) \in \left(1\pm\mu\right)\beta_{rms}^2\left(\{\alpha_{m,p}^{\left(0\right)}\}\right)
\label{eq:Problem2}
\end{multline}  
where $\bigl\| \cdot \bigr\|_{F}^2$ is the Frobenius norm squared and $\mathbf{R}\left\{\alpha_{m,p}\right\}$ is the MTSFM waveform set's correlation matrix.  As with the previous example, \eqref{eq:Problem2} is solved using Matlab's optimization toolbox \cite{Matlab}.  

\begin{figure}[ht]
\centering
\ifthenelse {\boolean{singleColumn}}
 {\includegraphics[width=1.0\textwidth]{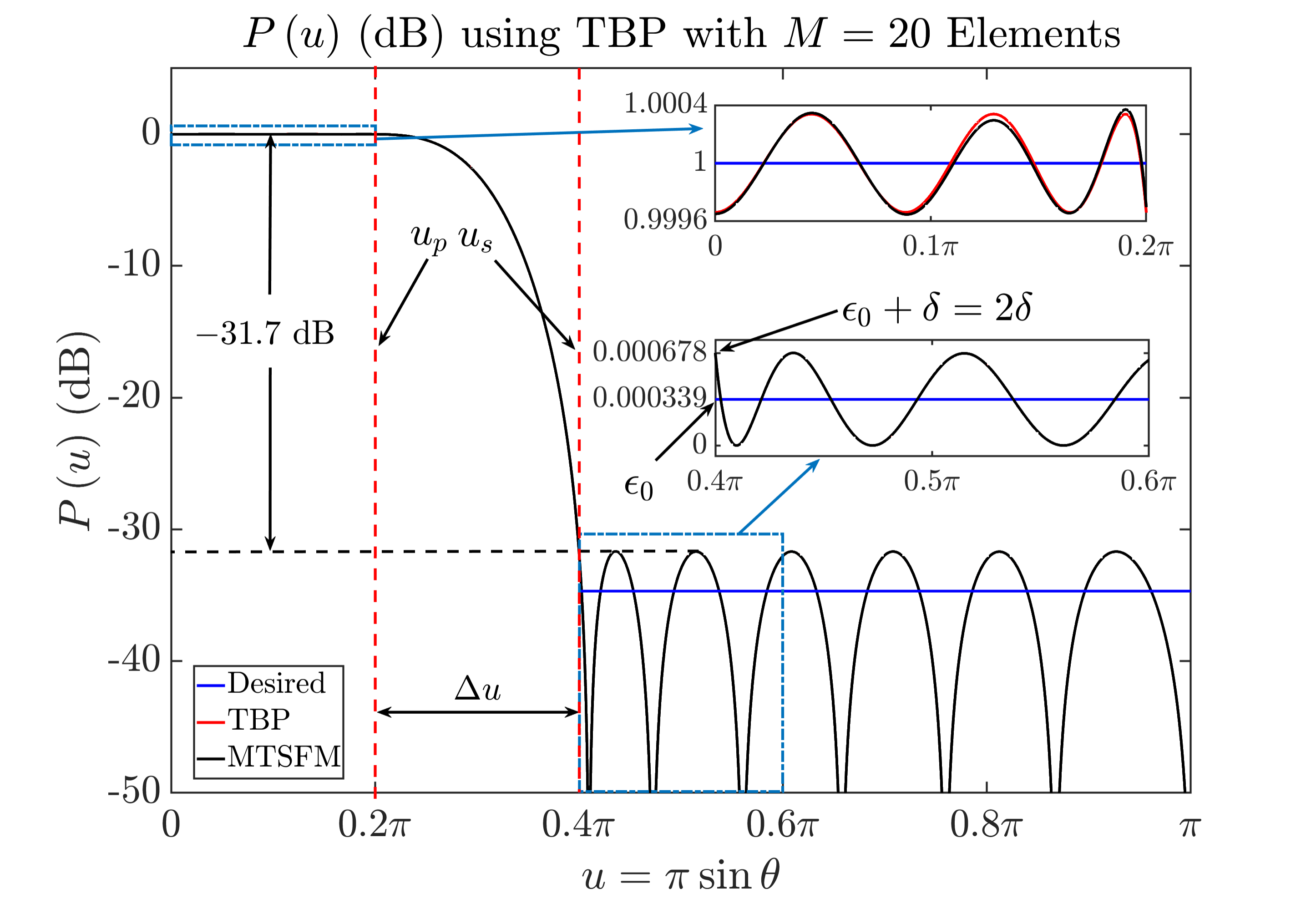}}
 {\includegraphics[width=0.5\textwidth]{Figure3.pdf}}
\caption{Equiripple MIMO beampatterns using the TBP method and orthogonal MTSFM waveform implementation.  There is only a slight deviation from the ideal TBP implementation in the passband of the MTSFM beampattern.  This is due to the MTSFM waveform set not being perfectly orthogonal.}
\label{fig:SPS_Figure_3}
\end{figure}

Figure \ref{fig:SPS_Figure_3} shows the resulting equiripple beampattern from direct implementation of the TBP technique and its MTSFM waveform set implementation.  The ripple in both the passband and stopband is constant as expected with only a small deviation in the passband of the MTSFM's beampattern.  This minute deviation is due to the off-diagonal elements of the MTSFM's correlation matrix.  While these elements were roughly 4 orders of magnitude smaller than the main diagonal elements, they are still non-zero and contributed a small perturbation to the resulting beampattern.  

\section{Conclusion}
\label{sec:Conclusion}
This letter described an equiripple MIMO beampattern synthesis method using Chebyshev approximation.  This is achieved by representing an even-symmetric MIMO beampattern as a nonnegative trigonometric polynomial which is identical to a Type-I FIR filter's frequency response.  The polynomial's coefficients are found efficiently using the Parks-McClellan algorithm.  These coefficients are then used to synthesize correlation matrices and constant envelope/spectrally compact waveform sets that realize the equiripple beampattern and was demonstrated with two design examples.  Future efforts will extend this model to incorporate a more general symmetry in the beampattern as was done with the frequency responses of equiripple FIR filters.


%


\ifCLASSOPTIONcaptionsoff
  \newpage
\fi

\clearpage
\section*{Acknowledgment}
D. A. Hague would like to acknowledge John R. Buck from the University of Massachusetts Dartmouth for his insightful conversations that inspired this work.

\end{document}